\documentclass[%
 reprint,
superscriptaddress,
 amsmath,amssymb,
prapplied,
aps,
]{revtex4-2}

\usepackage{graphicx}
\usepackage{dcolumn}
\usepackage{bm}
\usepackage{braket}
\usepackage{hyperref}
\hypersetup{
    colorlinks=true,
    linkcolor=blue,     
    urlcolor=blue,
    citecolor=blue,
}
\usepackage[capitalise]{cleveref}
\usepackage[euler]{textgreek}
\usepackage{xcolor}
\usepackage[version=4]{mhchem}
\usepackage{physics}
\usepackage{float}

\usepackage{subfigure}

\begin{document}

\preprint{APS/123-QED}


\title{Demonstration of Variational Quantum Eigensolver with Solid-State Spin System under Ambient Conditions}

\author{Xuliang Du}
\thanks{These three authors contributed equally}
\affiliation{%
 Department of Physics, The Hong Kong University of Science and Technology,
Clear Water Bay, Kowloon, Hong Kong, China
}%
\author{Yang Shen}
\thanks{These three authors contributed equally}
\affiliation{%
 Department of Physics, The Hong Kong University of Science and Technology,
Clear Water Bay, Kowloon, Hong Kong, China
}%
\author{Zipeng Wu}
\thanks{These three authors contributed equally}
\affiliation{%
 Department of Physics, The Hong Kong University of Science and Technology,
Clear Water Bay, Kowloon, Hong Kong, China
}%
\author{Bei Zeng}
\email{zengb@ust.hk}
\affiliation{%
 Department of Physics, The Hong Kong University of Science and Technology,
Clear Water Bay, Kowloon, Hong Kong, China
}%
\author{Sen Yang}
\email{phsyang@ust.hk}
\affiliation{%
 Department of Physics, The Hong Kong University of Science and Technology,
Clear Water Bay, Kowloon, Hong Kong, China
}


\begin{abstract}
Quantum simulators offer the potential to utilize the quantum nature of a physical system to study another physical system. In contrast to conventional simulation, which experiences an exponential increase in computational complexity, quantum simulation costs only linearly with the increasing size of the problem, rendering it a promising tool for applications in quantum chemistry.  The Variational Quantum Eigensolver (VQE) algorithm is a particularly promising application for investigating molecular electronic structures. For its experimental implementation, spin-based solid-state qubits have the advantage of long decoherence time and high-fidelity quantum gates, which can lead to high accuracy in the ground state finding. This study employs the nitrogen-vacancy (NV) center system in diamond to implement the VQE algorithm and successfully finds the eigenvalue of a specific Hamiltonian without the need for error mitigation techniques. With a fidelity of 98.9\% between the converged state and the ideal eigenstate, the demonstration provides an important step toward realizing a scalable quantum simulator in solid-state spin systems.
\end{abstract}

\maketitle

\section{Introduction}
Quantum computing has demonstrated its superiority over classical computing in numerous problem domains \cite{shor1994algorithms,grover1997quantum}, especially in the field of quantum simulation for the study of other physical quantum systems\cite{cao2019quantum}. However, the unavoidable noise from the environment leads to the decoherence of qubits and faulty quantum gates, which restricts further progress in the number of qubits and quantum circuit depth. To counter this, researchers have proposed theories such as quantum error correction to protect quantum systems \cite{nielsen2010quantum}, but their implementation requires a larger number of physical qubits, which in turn introduces more noise \cite{cory1998experimental,google2023suppressing}. 

Despite the current limitations on the number of qubits available, there are still various algorithms \cite{bharti2022noisy} that are capable of leveraging the advantage of quantum systems in this era of so-called noisy-intermediate-scale quantum (NISQ) devices \cite{preskill2018quantum}. Hybrid quantum-classical algorithms \cite{endo2021hybrid} are considered to be promising near-term applications, as they leverage quantum resources to complete classically consumable tasks while leaving other workloads on classical computing resources. The quantum parts of these algorithms often involve variationally updating a parameterized quantum circuit \cite{cerezo2021variational},
such as variational quantum eigensolver (VQE) \cite{peruzzo2014variational,mcclean2016theory,tilly2022variational,fedorov2022vqe} and quantum imaginary time evolution (QITE) \cite{motta2020determining,mcardle2019variational} for the ground state problems and quantum approximate optimization algorithm (QAOA)\cite{farhi2014quantum}. 

The VQE is a promising application of quantum simulation in the study of the electronic structures of molecules in quantum chemistry \cite{mcardle2020quantum}. The energy properties of the molecule are closely related to the ground state of its Hamiltonian. Conventionally, the computational complexity of finding the exact solution of the Hamiltonian increases exponentially with the size of the system. Nonetheless, since qubits evolve by the same nature of the quantum mechanics as the electronic wave functions, quantum simulation, as proposed by Feynman \cite{feynman2018simulating}, can be only linearly costly \cite{berry2007efficient} by simulating a quantum system by another quantum system. This feature makes VQE a useful tool for computing an upper bound of the Hamiltonian under study.

The experimental implementation of the VQE algorithm presents various challenges due to the requirements for expressibility and entangling capability\cite{sim2019expressibility}, which necessitate a certain circuit depth. Achieving this circuit depth typically demands a long decoherence time, posing a challenge for experimental realization. Furthermore, the iteration and gradient-based optimization process involved in VQE can be computationally intensive, further adding to the complexity of the experimental implementation.  The first experimental implementation of VQE is based on a photonic quantum processor \cite{peruzzo2014variational}. After that, VQE based on ion trap \cite{hempel2018quantum,nam2020ground,zhao2023orbital}, superconducting \cite{o2016scalable,kandala2017hardware,mccaskey2019quantum,google2020hartree,stanisic2022observing,huang2022simulating} and NMR \cite{hou2021spinq} systems are also realized. Compared with these systems, spin-based solid-state qubits such as the nitrogen-vacancy (NV) center spin in diamond offer advantages such as long decoherence time, high-fidelity quantum gates, and less stringent environmental requirements. However, the suitable ansatz for state preparation and the appropriate methods for state readout and the iteration strategy remain elusive for the implementation of the VQE algorithm on the NV center system.

In this work, we utilize the NV center spin in diamond and the nearest nitrogen nuclear spin to construct the quantum simulator. Although the nitrogen spin is normally a dark spin insensitive to the laser illumination, we realize the two-qubit read-out using fluorescence count difference at the excited state level anticrossing (esLAC). We choose the hardware-efficient ansatz for generating the trial state. Without error mitigation techniques, precise energy estimation is realized by Pauli term measurements with post-rotations after the state preparations. The minimization process to the eigenvalue is observed as the iteration times increase by optimizing the parameters with the gradient descent method. The process closely aligns with the predictions of numerical simulation.  The expectation value converges to -2.1 compared with the ideal eigenvalue -$\sqrt{5}$ (2.236). A fidelity of 98.9\% is achieved between the ideal eigenstate and the trial state represented by the converged parameters.

\section{Variational Quantum Eigensolver Algorithm}
A variational algorithm typically involves varying a set of parameters to minimize a defined loss function. For the VQE algorithm, the parameters $\vec{\theta}$ are for the quantum circuit to prepare the state $\ket{\psi(\vec{\theta})}$, which represents an approximation of the ground state of the certain Hamiltonian. The structure of the quantum circuit is defined as the variational ansatz. An effective ansatz can not only produce a better approximation of the exact eigenstate but also consume fewer computing resources. 

\begin{figure}[htb]
\centering
\includegraphics[width=1\columnwidth]{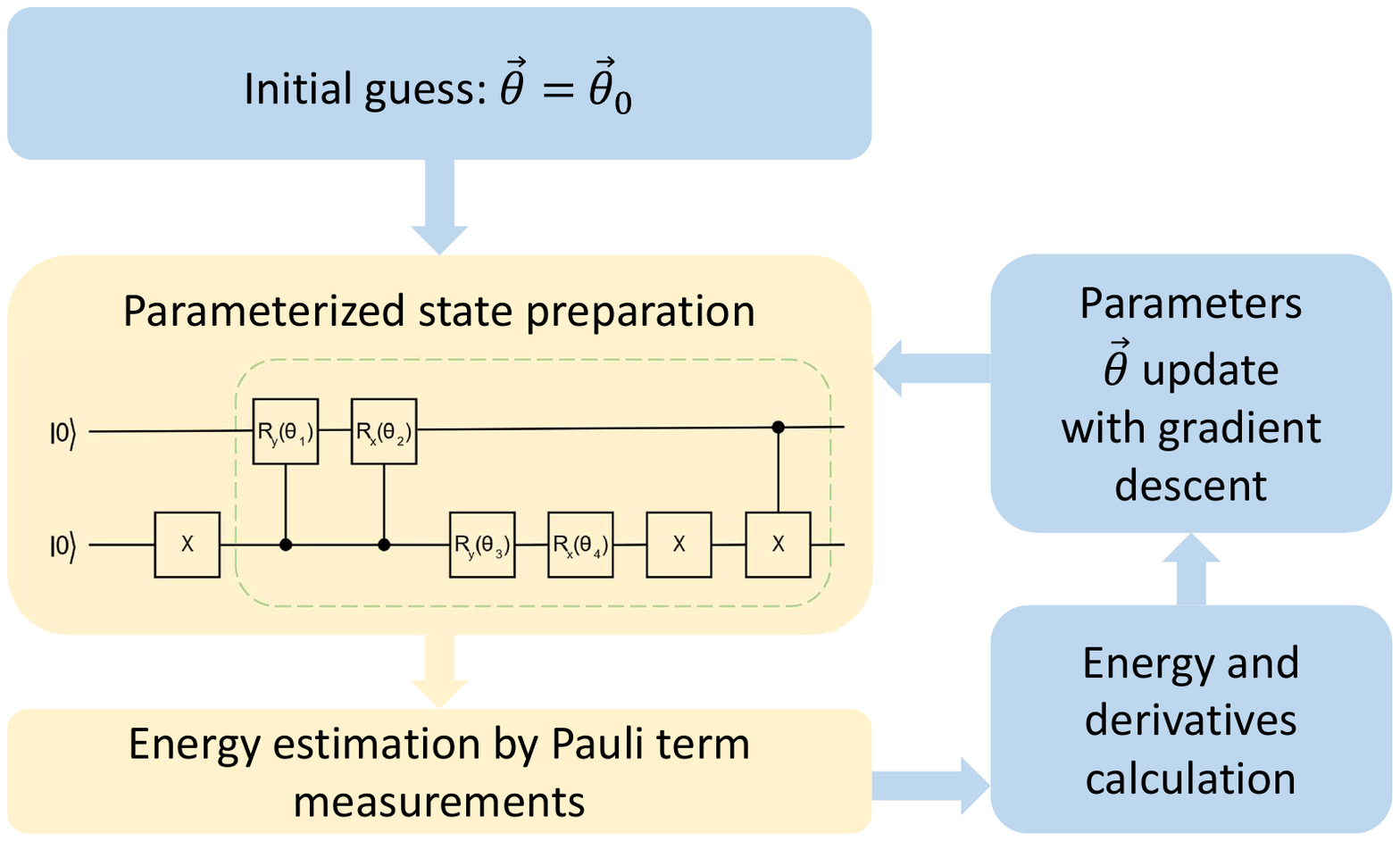}
\caption{\label{Procedure} The procedure of the VQE. The components in yellow color utilize the quantum qubits and the components in pale blue color are conducted in a classical computer. The qubit operations in the dashed box are defined as a unit layer in the current quantum circuit.}
\end{figure}

In order to investigate the lowest eigenvalue of the Hamiltonian, the loss function is naturally the expectation value
\begin{equation}
    E(\vec{\theta}) = \bra{\psi(\vec{\theta})} H \ket{\psi(\vec{\theta})}.
\end{equation}
For the minimization process of the expectation value, one can use the gradient-based optimizer in which the gradient is obtained by the parameter-shift rule \cite{schuld2019evaluating} \cite{mitarai2018quantum}
\begin{equation}
    \frac{\partial E(\vec{\theta})}{\partial \theta_i} = ( E(\vec{\theta}_i^+) - E(\vec{\theta}_i^-) )/ 2 
    \label{parameter-shift rule}
\end{equation}
where $\vec{\theta}_i^\pm = \vec{\theta} \pm \frac{\pi}{2} \vec{e}_i$, $\vec{e}_i$ is the $i$th unit vector in the parameter space. Thus the neighboring states in parameters space are described as $\ket{\psi(\vec{\theta}_i^\pm)}$. With the gradient, the parameters are updated using the gradient descent method
\begin{equation}
    \vec{\theta}^{\,'} = \vec{\theta} - \alpha \nabla E(\vec{\theta}) .
    \label{gradient descent}
\end{equation}

The algorithm distributes the workload between classical and quantum computing resources. It proceeds through a series of steps, as depicted in Fig. \ref{Procedure}. A typical iteration of the VQE starts with an initial guess of the parameters $\vec{\theta}$ before entering the loop. According to this set of parameters, a quantum circuit is then employed to apply the unitary operation $U(\vec{\theta})$ to the initial state, generating the trial state $\ket{\psi(\vec{\theta})}$. To gain the gradient information about this set of parameters, states that are nearby in parameter space are also generated. The states are subsequently measured, and the information about the expectation value of the Hamiltonian is transferred to classical computing resources. The expectation values $\bra{\psi(\vec{\theta})} H \ket{\psi(\vec{\theta})}$ of these states can then be calculated. From these expectation values, the gradients of this set of parameters can also be calculated. The parameters are updated using the gradient descent method, and a new iteration begins with the updated parameters. The loop continues until the expectation value converges to the minimum.

\section{Experimental Implementation}

\begin{figure}[htb]
\centering
\includegraphics[width=0.8\columnwidth]{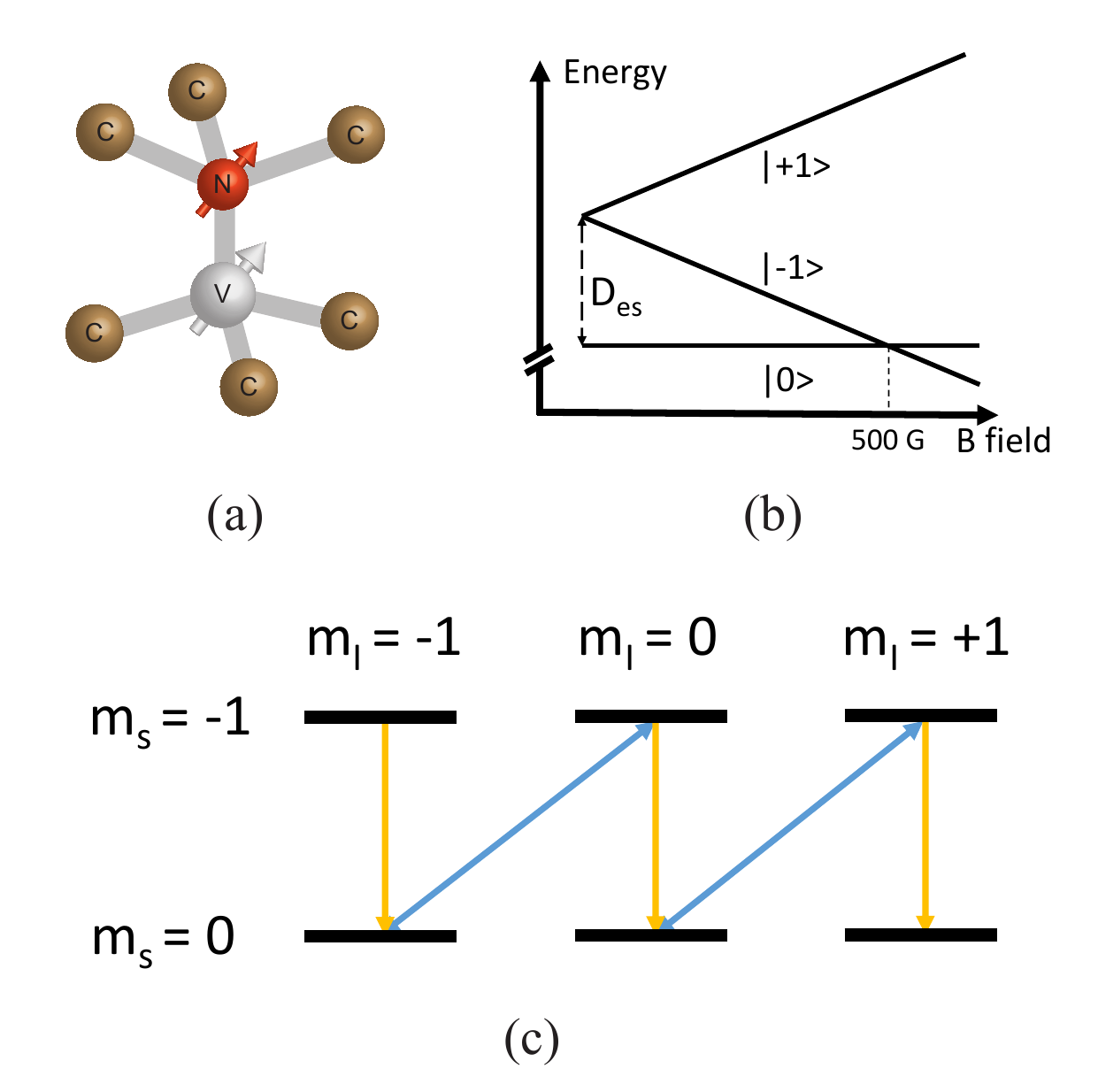}
\caption{ Experimental system for VQE. (a) A schematic illustration of the NV center spin and nitrogen nuclear spin as qubits. (b) Spin energy levels in excited states for various magnetic fields. $\ket{m_s=0}$ and $\ket{m_s=-1}$ states cross at around 500~Gauss, causing esLAC. (c) Optical initialization at the esLAC. The strong hyperfine coupling between the electron and the nuclear spin enables the energy-conserving flip-flop process between them. Yellow single-headed arrows mean laser initialization of the NV spin to the $\ket{m_s=0}$ states and purple double-headed arrows characterize the flip-flop process.}
\label{energy structure}
\end{figure}

The VQE algorithm is implemented on a two-qubit system composed of the NV center spin and the nitrogen nuclear spin. The negatively charged NV center is a form of point defect in diamond that acts as a spin-one system \cite{doherty2011negatively,jelezko2006single}. The NV center has a long coherence time \cite{herbschleb2019ultra}, making it a favorable option for realizing a quantum simulator. The naturally occurring nitrogen nuclear spin of the NV can be coupled to it with hyperfine interaction, forming a two-qubit system. A schematic illustration of the two-qubit system of NV spin and the nitrogen nuclear spin is shown in Fig. \ref{energy structure}(a). The details of our experimental setup will be discussed in Appendix \ref{Experiment Setup}. Our implementation of the VQE algorithm utilizes the $m_s = 0$ and $m_s = -1$ subspace of the NV center and the $m_I = +1$ and $m_I = 0$ subspace of the nitrogen nuclear to define the qubits. $m_s$ is the magnetic quantum number of the electron spin and $m_I$ of the nitrogen nuclear spin. In this paper, we designate the nitrogen nuclear spin as qubit 1 with parameters $\theta_1$ and $\theta_2$ for y and x rotations as in quantum circuit Fig. \ref{Procedure}. Likewise, the NV center spin is identified as qubit 2, with parameters $\theta_3$ and $\theta_4$. We carefully selected the four parameters in a way that each spin possesses two degrees of freedom. This decision was made based on our prior knowledge of the Hamiltonian, allowing us to effectively represent its solution space while minimizing the number of gates required. We set the magnetic field around 500~Gauss to reach the excited state level anticrossing (esLAC), with which we can initialize the system by optical illumination \cite{jacques2009dynamic} as shown in Fig. \ref{energy structure}(b)(c). The esLAC-caused spin-dependent fluorescence is also the key to reading out the system \cite{van2012decoherence}. The coherent control of the NV electron spin and nitrogen nuclear spin is realized by applying microwave and radio-frequency pulses.
 
The Hamiltonian under study is 
\begin{equation}
    H = X_1 X_2 + Z_1 + Z_2.
\end{equation}
The minimum eigenvalue of this Hamiltonian should be $-\sqrt{5}$, and it corresponds to the eigenstate $ -\frac{\sqrt{50-20\sqrt{5}}}{10}\ket{00} + \frac{\sqrt{50+20\sqrt{5}}}{10}\ket{11}$. Though it is obvious that the straightforward Pauli measurements are $X_1 X_2$, $I_1 Z_2$ and $Z_1 I_2$, measurements of $I_1 I_2$, $I_1 Z_2$, $Z_1 I_2$, $Z_1 Z_2$, $X_1 X_2 + Y_1 Y_2$ and $X_1 X_2 - Y_1 Y_2$ are needed in our experiments due to the restriction of the NV center system, which will be discussed in detail in Appendix B  \ref{Appendix Pauli}. 

\begin{figure}[htb]
\centering
\includegraphics[width=1\columnwidth]{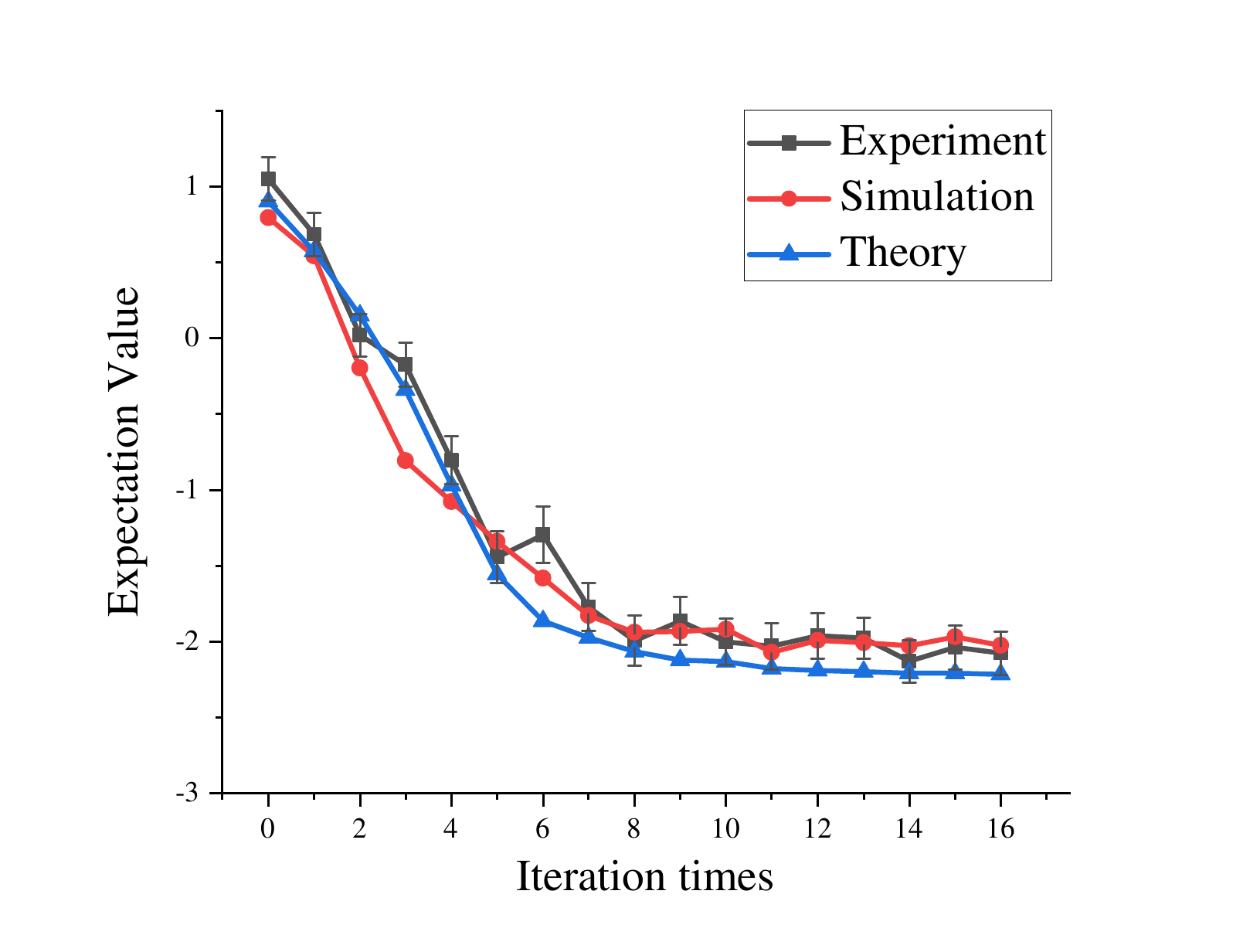}
\caption{\label{result} The minimization process of the expectation value during VQE iterations. The brown line represents the experiment result. The orange line represents the simulated minimization process with the same initial guess and learning rate. The specific parameters for each iteration in the simulation line are calculated according to the simulation result of the last iteration, different from the experiment parameters. The blue line represents the theoretical expectation value for the same set of parameters in the experiment of each iteration.}
\end{figure}

\begin{figure}[htb]
\centering
\includegraphics[width=0.8\columnwidth]{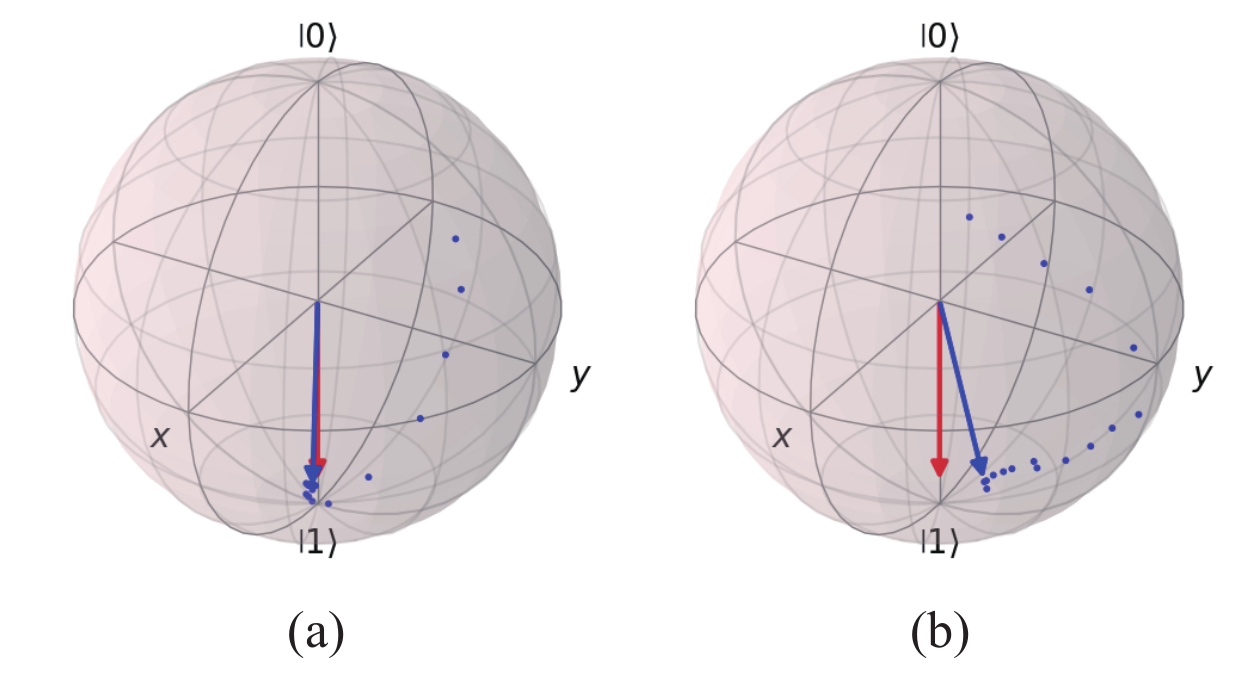}
\caption{ The minimization process in Bloch sphere of (a) the nitrogen nuclear spin qubit and (b) NV spin qubit.  The red arrow represents the ideal eigenstate and the blue arrow represents the state of the converged parameters. The blue dots represent the state of the parameters in different iterations.}
\label{bloch sphere}
\end{figure}

\begin{figure*}[htb]
\centering
\includegraphics[width=2\columnwidth]{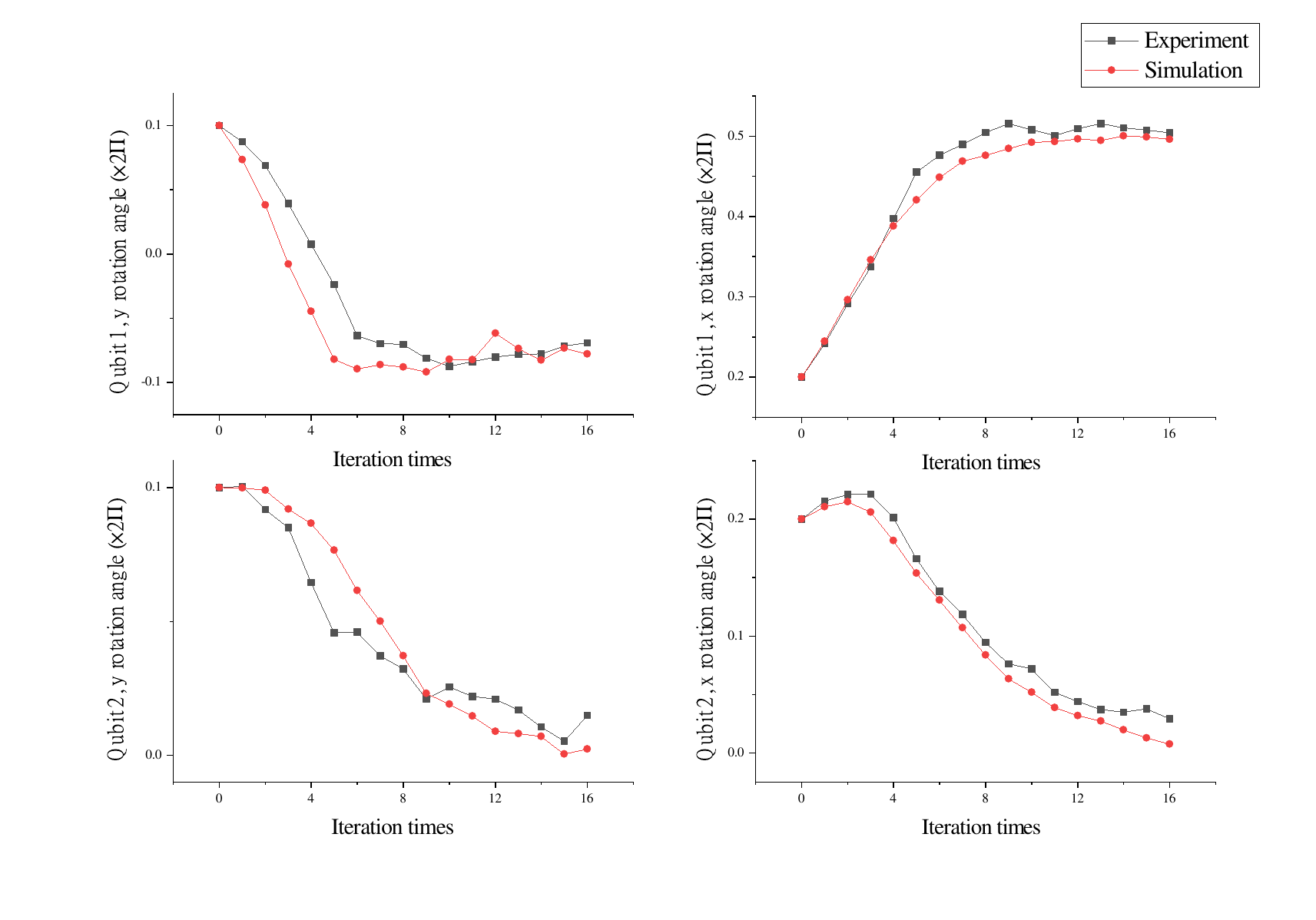}
    
\caption{The specific optimization processes for four specific parameters. Four figures correspond to the nuclear(qubit 1) and electron(qubit 2) spin rotation angles in the y and x directions respectively. The brown line represents the experiment result. The orange line represents the simulated minimization process with the same initial guess and learning rate.}
\label{four specific parameters}
\end{figure*}

The experimental procedure of the VQE can be summarized into these steps: 
\paragraph{Choose an initial guess of the parameters $\vec{\theta}_0$ and an appropriate learning rate $\alpha$.}
The four components of the $\vec{\theta}_0$ correspond respectively to the nuclear and electron spin rotation angles in the y and x directions. The learning rate $\alpha$ for the updating of the parameters can affect both the optimization process and the final expectation value the process converges to. An appropriate learning rate is expected to be small enough to cause the convergence minimum close to the ideal eigenvalue but won't make the convergence process too slow. The closeness also reaches a plateau when $\alpha$ is below a certain point.
\paragraph{Prepare and measure the quantum states.}
The trial state $\ket{\psi(\vec{\theta})}$ is prepared with the hardware-efficient ansatz depicted in Fig. \ref{Procedure}, in which the parameter $\theta_i$ representing the rotation angle is encoded as the duration of a driving pulse in hardware. In this paper, we use one layer in the state preparation to generate the trial state. By post-rotations after the state preparation, measurements of certain Pauli matrices are achieved. According to the specific Hamiltonian, we can choose the specific Pauli terms to measure. The preparation and the readout are also repeated for eight other states $\ket{\psi(\vec{\theta}_i^\pm)}$ neighboring in the parameter space. In the course of experimentation, it is imperative to execute a particular sequence for a given state repeatedly for a significant number of instances to ensure reliability and precision in the obtained results. In this regard, it is noteworthy that a million trials are conducted for each sequence corresponding to a single state in our experiment.

\paragraph{Estimate the expectation values and the gradients.}
The expectation values are calculated on a classical computer with the results of Pauli measurements on a quantum computer for the trial state and eight neighboring states. The four gradients of the expectation value in $\vec{\theta}_0$ are calculated by the parameter-shift rule [Eq. \ref{parameter-shift rule}] with the estimated expectation values of 8 neighboring states.
\paragraph{Update the parameters.}
The parameters are updated using the gradient descent method [Eq. \ref{gradient descent}] with the four gradients calculated in the last step. 
\paragraph{Repeat steps (b) (c) (d) until the expectation value converges to the minimum.}
~\\

We start the VQE experiment with a random initial guess of the parameters as $\vec{\theta}_0 = (0.1, 0.2, 0.1, 0.2) \times 2\pi $. In order to reduce the duration of the experiment without compromising the optimization intricacies, the value of the learning rate is set to $\alpha = 0.3$. The algorithm converges to $E(\vec{\theta}) \approx -2.1$ after 16 iterations as shown in Fig. \ref{result} as the brown line. The trial state represented by the converged parameters has a fidelity of 98.9\% with the ideal eigenstate. The trajectories in the Bloch sphere of the minimization process for two spins are depicted as the blue dots in Fig. \ref{bloch sphere}. The dots in the Bloch spheres represent the reduced density matrices of two spins. For the specific optimization processes of four parameters, experiment and simulation results are included in Fig. \ref{four specific parameters}. From the optimization processes of the parameters of the electron spin qubit, we can infer more iterations may eliminate the distance between the converged state and the ideal eigenstate of the NV spin qubit in Fig. \ref{bloch sphere}.


By comparing with the simulations in Qiskit\cite{Qiskit}, we find that the method based on Pauli term measurements to estimate expectation value has a read-out error at the magnitude of approximately 0.01, which is discussed in detail in Appendix \ref{Appendix readout error}. The major noise contributing to the fluctuation of the measured expectation value is the photon shot noise. Although the percentage error of the fluorescence count reaches 0.1\% as a result of a million repetitions for each state, the fluctuation of the final calculated value is larger due to the error propagation. Besides, the inhomogeneity of the magnetic field in different runs of the experiment and the error of the fitted resonance frequency affect not only the precision of the estimation of the expectation value but also the generation of the desired state, causing the difference between the VQE-optimized expectation value and the ideal eigenvalue. After estimating the magnitude of the dephasing process through $T_2^*$ experiments, we add the above noises and simulate the experiment in Lindblad equation with Qutip \cite{JOHANSSON20121760, JOHANSSON20131234} using the same initial guess of the parameters, learning rate and other experimental parameters, which is also shown in Fig. \ref{result}. The estimation of the error bar in the experimental result is accomplished through a method inspired by the Monte Carlo experiment. More specifically, the method involves the random sampling of direct measurement outcome errors, which are based on their corresponding experimental photon shot noises. Subsequently, the error-added results go through normal data processing, resulting in the generation of a distribution for the expectation value. The error bar is then defined as the standard deviation of this distribution.

\section{conclusion}
We demonstrate the variational quantum eigensolver algorithm on a two-qubit system based on the nitrogen-vacancy center in diamond under ambient conditions. A decay to the lowest eigenvalue for the Hamiltonian under study is observed as predicted by the numerical simulation with our noise model. The accuracy can be further increased by using optimal control \cite{rong2015experimental} and error mitigation techniques.

Although only a simple VQE algorithm with four parameters is implemented in this paper, the simulation power can be broadened by leveraging all three levels of the NV center spin and nitrogen nuclear spin to construct a quantum simulator of two qutrits.
Except for VQE, other quantum-classical algorithms like QAOA can also benefit from this qutrit system \cite{bottrill2023exploring}. The initialization and read-out methodology can be generalized from the existing technique based on esLAC. 
Besides, more layers in the parameterized state preparation can be added to enhance the expressibility and entangling capability \cite{sim2019expressibility}.

Furthermore, the demonstration has substantial scalability and versatility for extension to other fields of solid-spin systems. To scale up the system, besides using nuclear spins qubits around the color center \cite{van2024mapping,shen2023detection}, we can also develop a connected network of entangled nodes, each containing a limited number of qubits \cite{awschalom2018quantum}. The coupling between nodes has been realized on NV centers \cite{bernien2013heralded} and group-IV vacancies in diamond \cite{sipahigil2016integrated}. This can help the VQE to study more complex Hamiltonians or harness parallelism to accelerate the optimization process \cite{mineh2023accelerating}. Moreover, we demonstrate the feasibility of implementing the VQE on hybrid spin systems, where coherent control and readout present challenges due to the distinct spin species and one spin functioning as a readout probe. Our approach to this issue and the entangled nodes architecture can be extended to other solid-state hybrid spin platforms with the similar problem, such as group-IV vacancies in diamond, rare-earth ions in solids \cite{kinos2021roadmap}, and divacancies in silicon carbide \cite{bourassa2020entanglement}. Besides, our Pauli measurements technique is essentially an efficient 2-qubit tomography method, which can be adapted into other quantum applications such as quantum sensing.

\begin{acknowledgments}
We thank Cheuk Kit Cheung for the fruitful discussion. B.Z. acknowledges
financial support from Hong Kong RGC (GRF/16305121). S.Y. acknowledges
financial support from Hong Kong RGC (GRF/16305422).

Xuliang Du, Yang Shen, and Zipeng Wu contributed equally to this work.
\end{acknowledgments}

\appendix

\section{Experiment Setup}
\label{Experiment Setup}
\begin{figure*}[htb]
\centering
\includegraphics[width=2\columnwidth]{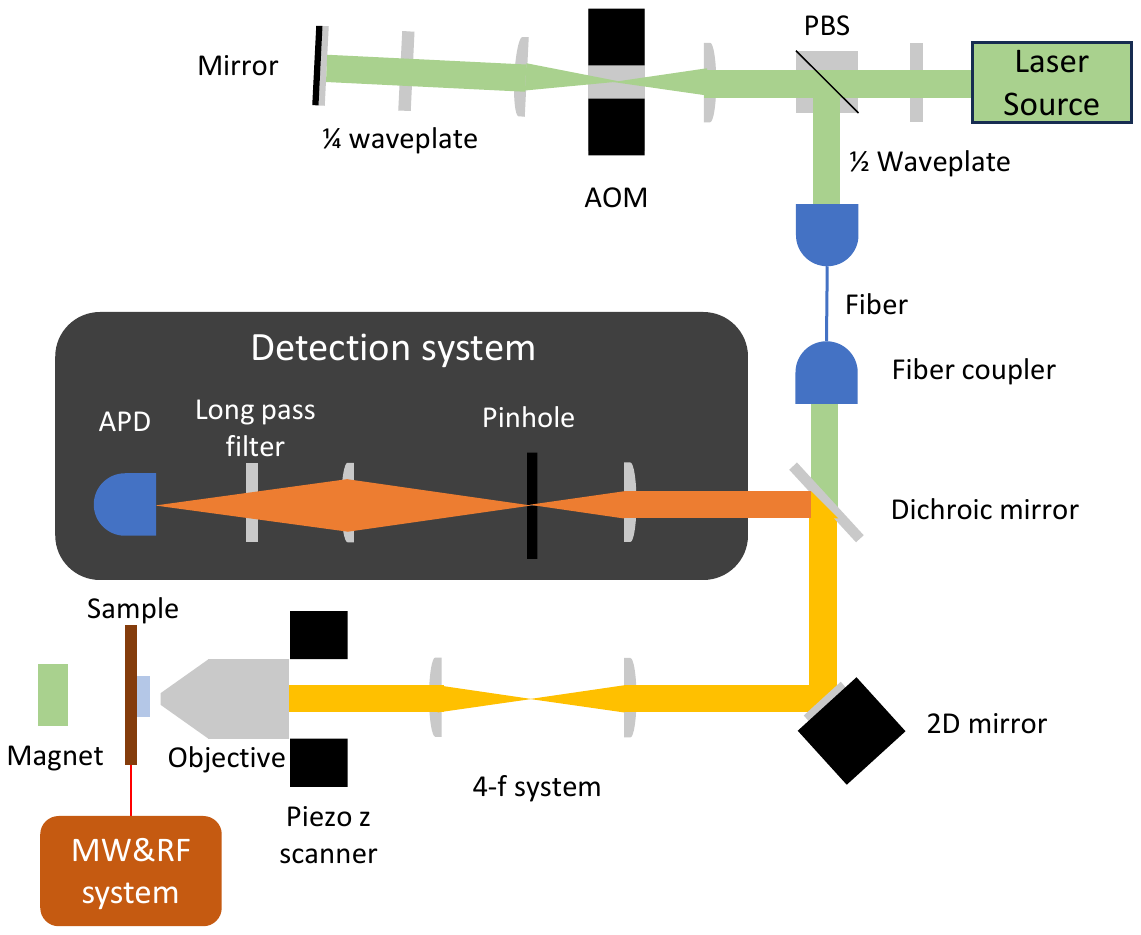}
\caption{\textbf{Experiment setup} }
\label{Experiment setup}
\end{figure*}
The experimental configuration is illustrated in Figure \ref{Experiment setup}. We use a type \uppercase\expandafter{\romannumeral2}a natural diamond sample oriented along the 111 crystallographic direction with a 1.1\% natural abundance of $^{13} C$. The diamond sample is affixed to a sample holder, which has a copper wire located near the NV center and is connected to the microwave and radio-frequency system. To enhance the collection efficiency, a solid immersion lens (SIL) is fabricated around the NV center using a focused ion beam. Additionally, a permanent magnet is situated in proximity to the sample to create a magnetic field parallel to the NV axis.

The green laser utilized for NV initialization is generated using a 530~nm laser source (Coherent Sapphire) and is subsequently directed through a half-waveplate to modify the polarization direction. Then it first passes the polarization beam splitter(PBS). After double passing the acousto-optic modulator(AOM, Gooch\&Housego 3350-199) and a quarter waveplate altering the polarization, the laser got reflected by the PBS with a perpendicular polarization to the original one and coupled into a fiber. The AOM, which can function as a switch for the green laser, is controlled by a 350~MHz microwave. After exiting the fiber, the green laser passes through a dichroic mirror and is reflected by a piezo-controlled 2D mirror before entering the 4f system, which includes two additional lenses. The 2D mirror enables the xy-plane scan of the diamond sample. Following the second lens, the green light is directed into an oil 100$\times$ objective (UPLSAPO100XO) with a numerical aperture of 1.4 mounted on the piezo z scanner. Finally, the laser reaches the NV center, and the z scanner controls the objective's z position, enabling the z-direction scan of the diamond sample.

The red fluorescence emitted by the NV center propagates back along the same optical path as the green laser until it reaches the dichroic mirror, where it is reflected toward the detection system. The lens and the pinhole work like a spatial filter of fluorescence. After the pinhole, a long pass filter is used to further filter out the non-fluorescence photon. In the end, the fluorescence is focused into an avalanche photon detector(APD, SPCM-AQRH-10-FC, Excelitas Technologies).

\section{Pauli measurements}
\label{Appendix Pauli}
\begin{figure}[htb]
\centering
\includegraphics[width=0.8\columnwidth]{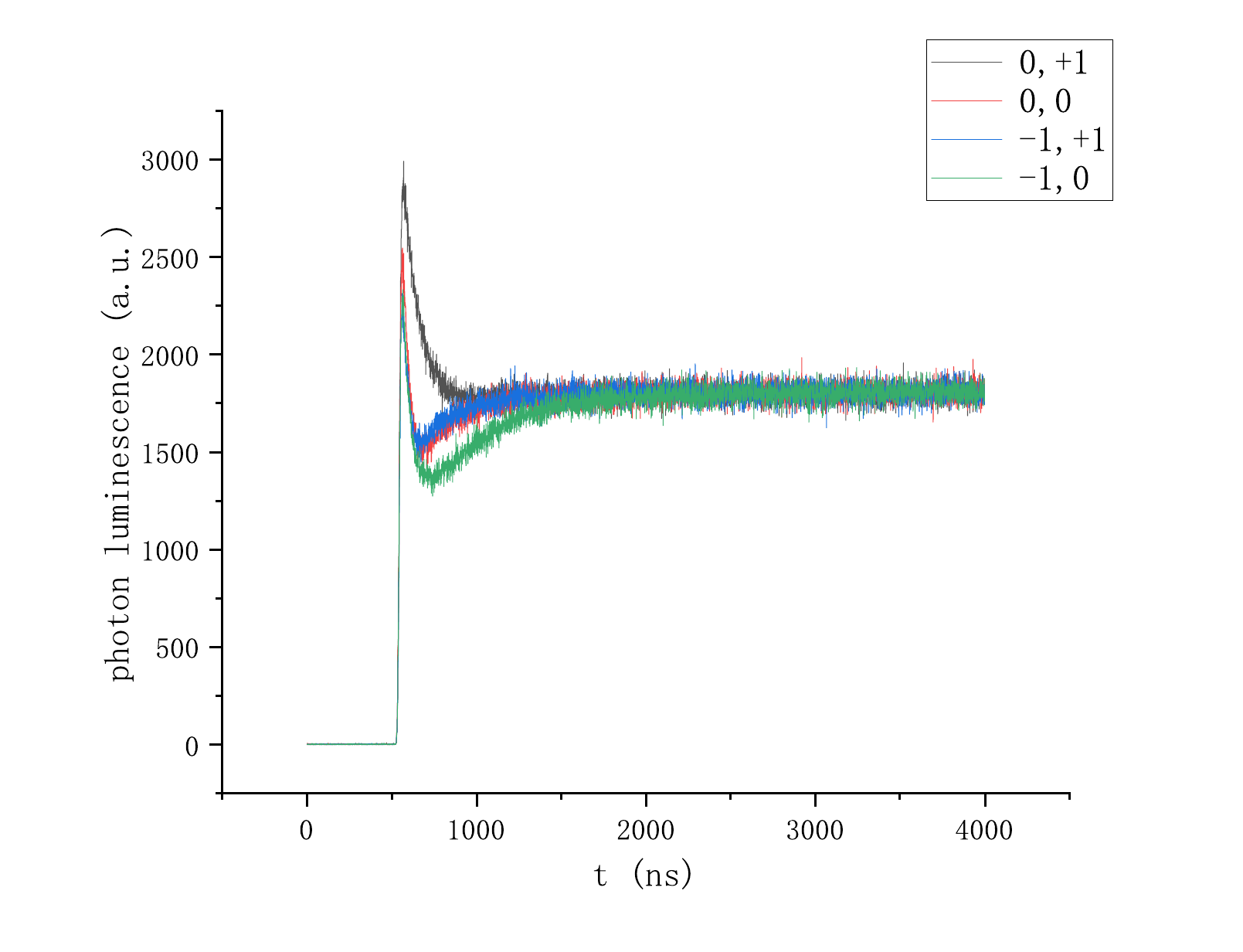}
\caption{\label{PL} A typical time-binned photon luminescence of different state}
\end{figure}
Pauli measurements after the trial state preparation are based on the different photoluminescence (PL) rates of different states under the condition of the excited state level anticrossing (esLAC)\cite{steiner2010universal,van2012decoherence}. Fig. \ref{PL} shows the typical time-binned PL of the NV-nitrogen two-qubit system in different states. Since the photoluminescence of an arbitrary state simply depends on its population in different states, the measurement matrix without post-rotation is simply
\begin{equation}
    M_{0} = 
    \begin{pmatrix}
        N_1 & 0 & 0 & 0 \\
        0 & N_2 & 0 & 0 \\
        0 & 0 & N_3 & 0 \\
        0 & 0 & 0 & N_4 \\
    \end{pmatrix}
\end{equation}
in which N is the PL count and $1, 2, 3, 4$ denote the states $\ket{0,+1}$,$\ket{0,0}$,$\ket{-1,+1}$,$\ket{-1,0}$ respectively.

To realize Pauli measurements $I_1 I_2$, $I_1 Z_2$, $Z_1 I_2$, $Z_1 Z_2$, we also do the measurements with the post rotation of $\pi$ pulses between these states. By doing this, we realize the measurement matrices
\begin{equation}
    M_{\pi_{13}} = 
    \begin{pmatrix}
        N_3 & 0 & 0 & 0 \\
        0 & N_2 & 0 & 0 \\
        0 & 0 & N_1 & 0 \\
        0 & 0 & 0 & N_4 \\
    \end{pmatrix}
\end{equation}

\begin{equation}
    M_{\pi_{34}} = 
    \begin{pmatrix}
        N_1 & 0 & 0 & 0 \\
        0 & N_2 & 0 & 0 \\
        0 & 0 & N_4 & 0 \\
        0 & 0 & 0 & N_3 \\
    \end{pmatrix}
\end{equation}

\begin{equation}
    M_{\Pi} = 
    \begin{pmatrix}
        N_3 & 0 & 0 & 0 \\
        0 & N_4 & 0 & 0 \\
        0 & 0 & N_1 & 0 \\
        0 & 0 & 0 & N_2 \\
    \end{pmatrix}
\end{equation}
in which $M_{\pi_{13}}$ means a $\pi$ pulse between states $\ket{0,+1}$ and $\ket{-1,+1}$ is applied after the trial state preparation and before the photon count measurement. $M_{\pi_{34}}$ corresponds to a $\pi$ pulse between states $\ket{-1,0}$ and $\ket{-1,+1}$ and $M_{\Pi}$ corresponds to a hard $\pi$ pulse which flip both $\ket{0,+1} \longleftrightarrow \ket{-1,+1}$ and $\ket{0,0} \longleftrightarrow \ket{-1,0}$.

With the result of $M_0,M_{\pi_{13}},M_{\pi_{34}},M_{\Pi}$ as $R_0,R_{\pi_{13}},R_{\pi_{34}},R_{\Pi}$, we can get the Pauli measurement results 
\begin{widetext}
\begin{equation}
    \begin{pmatrix}
        R_{I_1 I_2} \\
        R_{I_1 Z_2} \\
        R_{Z_1 I_2} \\
        R_{Z_1 Z_2} \\
    \end{pmatrix} = 
    \begin{pmatrix}
        1 & 1 & 1 & 1 \\
        1 & -1 & 1 & -1 \\
        1 & 1 & -1 & -1 \\
        1 & -1 & -1 & 1 \\
    \end{pmatrix}
    \begin{pmatrix}
        N_{1} & N_{2} & N_{3} & N_{4} \\
        N_{3} & N_{2} & N_{1} & N_{4} \\
        N_{1} & N_{2} & N_{4} & N_{3} \\
        N_{3} & N_{4} & N_{1} & N_{2} \\
    \end{pmatrix}^{-1}
    \begin{pmatrix}
        R_0 \\
        R_{\pi_{13}} \\
        R_{\pi_{34}} \\
        R_{\Pi} \\
    \end{pmatrix}.
\end{equation}
\end{widetext}

\begin{figure*}[htb]
\centering
\includegraphics[width=2\columnwidth]{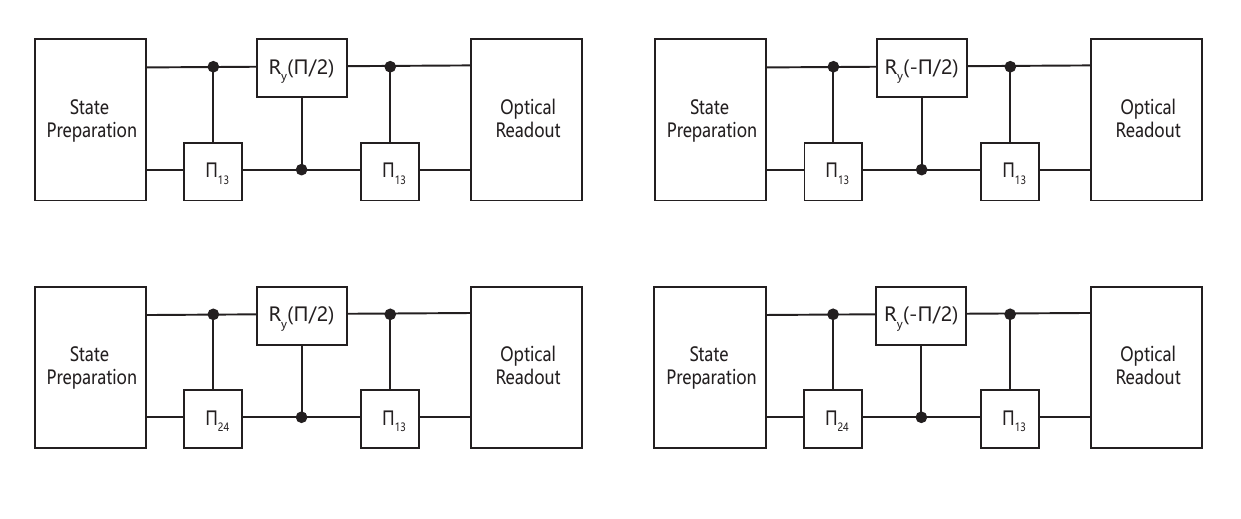}
\caption{\label{qc meas} Quantum circuits of the post-rotations for the measurement of $X_1 X_2 + Y_1 Y_2$ and $X_1 X_2 - Y_1 Y_2$.}
\end{figure*}

To realize the Pauli measurement of $X_1 X_2$, we have to do the measurement of both $X_1 X_2 + Y_1 Y_2$ and $X_1 X_2 - Y_1 Y_2$. The $X_1 X_2 + Y_1 Y_2$ term is measured by considering the PL difference between two sequences with post-rotations. The difference is normalized by the difference between the states $\ket{0,+1}$ and $\ket{-1,0}$. The first sequence consists of a $\pi$ pulse between the states $\ket{0,+1}$ and $\ket{-1,+1}$, followed by a $\frac{\pi}{2}$ pulse around the y-axis of the nitrogen nuclear spin in the rotating frame between states $\ket{-1,+1}$ and $\ket{-1,0}$, and finally another $\pi$ pulse between states $\ket{0,+1}$ and $\ket{-1,+1}$. The second sequence is identical, except for an opposite rotation around the -y axis for the second pulse. The $X_1 X_2 - Y_1 Y_2$ term is measured the same way but the first pulse for the two sequences is a $\pi$ pulse between the states $\ket{0,0}$ and $\ket{-1,0}$. The quantum circuits of the post-rotations for the measurement of $X_1 X_2 + Y_1 Y_2$ and $X_1 X_2 - Y_1 Y_2$ are shown in Fig. \ref{qc meas}.

\section{Phase consideration in quantum circuit}
Owing to the limited power of the microwave, it becomes crucial to account for the unavoidable duration time of the quantum gate imposed on the electron spin when conducting the in-phase measurement of the nuclear spin. Throughout the operation of the quantum gate, the electron spin is in a superposition state of $\ket{0}$ and $\ket{-1}$. Given the intimate connection between the nuclear and electron spins, facilitated primarily by the hyperfine interaction, differing states of the electron spin can prompt disparate energy splittings between the up and down states of the nuclear spin. Consequently, the nuclear spin undergoes precession around the direction of the magnetic field with a frequency that is contingent on the state of the electron spin. For the purpose of manipulating the nuclear spin within the rotating frame, one must contemplate the phase discrepancy induced by this disparity in frequency.

All radio-frequency (rf) pulses employed for the manipulation of the nuclear spin are in resonance with the transition between the $\ket{-1,+1}$ and $\ket{-1,0}$ states. Hence, we opt to probe the nuclear spin in this particular rotating frame. Nevertheless, the parameters $\theta_3$ and $\theta_4$ imply that a part of the electron spin's population will undergo rotation towards and back from the state $\ket{0}$ upon application of the microwave. Within the previously stated rotating frame, the nuclear spin will precess at a frequency of 2.16 MHz (the hyperfine interaction strength) when the electron spin is in the state $\ket{0}$. It thus follows that the phase of measurement gates for the nuclear spin, subsequent to the preparation of the trial state, requires modification after the process through the state $\ket{0}$. Besides, the measurement result also needs to be modified because of the phase spreading during the process. Through the simulation of the dynamics under the influence of the microwave directly using Schrodinger's equation with Qutip\cite{JOHANSSON20121760, JOHANSSON20131234}, we derive the phase discrepancy for the rf pulse in the post-rotation sequences and the correction factor for the readout results.

\section{Simulation}
A common method to describe the dynamic of a system interacting with the environment is the Lindblad master equation, which has the form
\begin{widetext}
    \begin{equation}
    \dot{\rho}(t) = - \frac{i}{\hbar}[H(t),\rho(t)] + \sum_n \frac{1}{2} [2 C_n \rho(t) C_n^\dagger - \rho(t) C_n^\dagger C_n - C_n^\dagger C_n\rho(t)].
\end{equation}
\end{widetext}

\begin{figure}[htb]
    \centering
    \includegraphics[width=0.8\columnwidth]{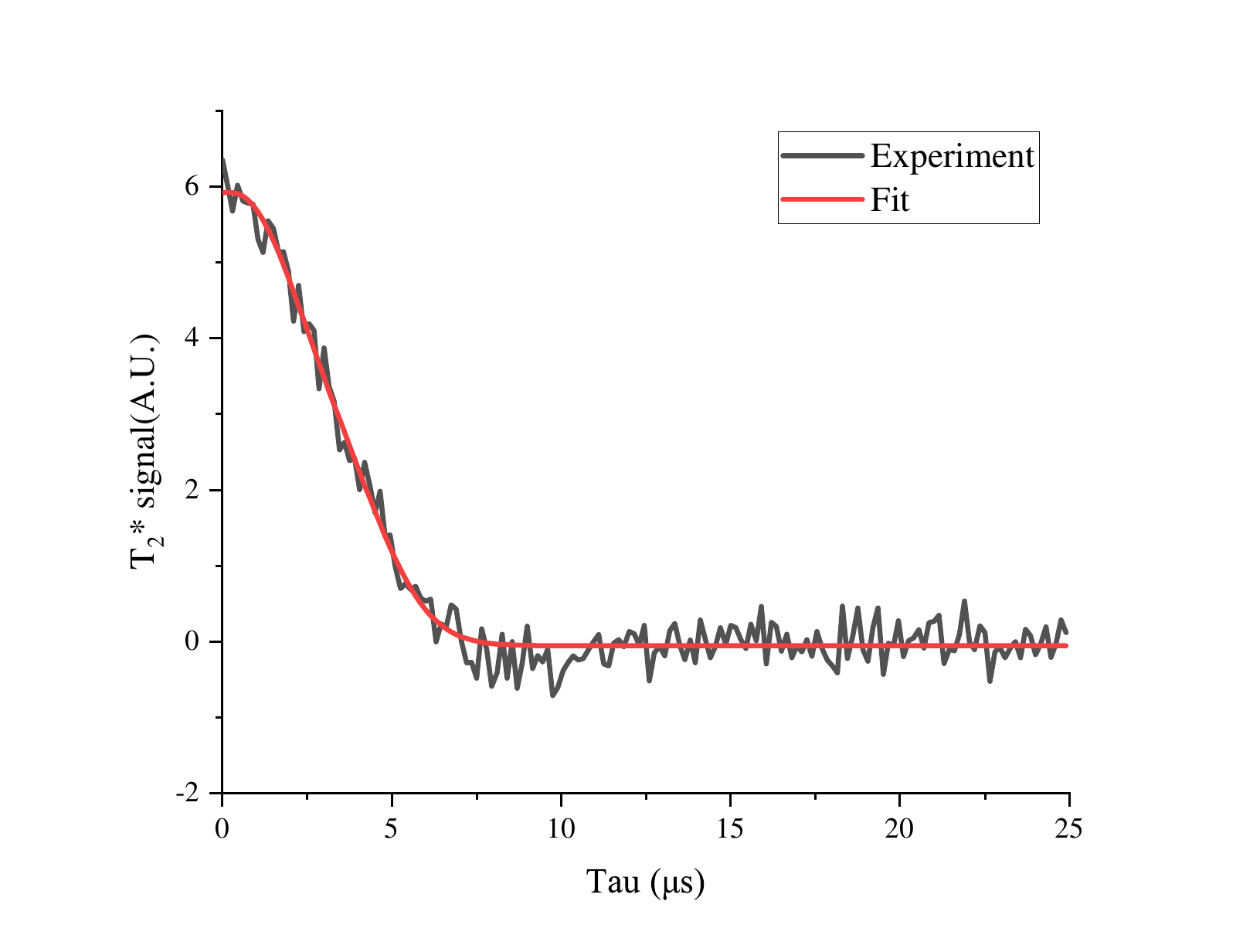}
    \caption{\label{T2star} $T_2^*$ measurement with experimental data and the fit.}
\end{figure}

Here, $C_n$ is the collapse operator characterizing the influence of the environment. Since the dephasing process caused by the inhomogeneity of the magnetic field plays a major role in decoherence in our experiment, we mainly consider the collapse operator with the form
\begin{equation}
    C_n =   \begin{pmatrix}
                0 & 0  \\
                0 & 1  \\
            \end{pmatrix}.
\end{equation}
Through the $T_2^*$ measurement shown in Fig. \ref{T2star}, we estimate the $T_2^*$ is around 4.367~$\mu s$ by which we can estimate the amplitude of the collapse operator. 

With the knowledge of the decoherence and specific experimental parameters, we next use the Lindblad equation in qutip to compute the dynamics of the two-qubit system under the microwave and radiofrequency driving in each experimental sequence. To account for photon shot noise, we introduce a white noise of the magnitude equivalent to the experimental signal.  The simulated outcomes then go through the same data processing procedures as the experimental data, yielding the simulated expectation value for a specific parameter set. In the simulation results, only the initial point shares the same parameters as the experiment, while all subsequent iteration points are generated based on the gradients calculated using the expectation value obtained during the simulation.

\section{Converged expectation value and read-out error}
\label{Appendix readout error}
\begin{figure}[htb]
\centering
\includegraphics[width=0.8\columnwidth]{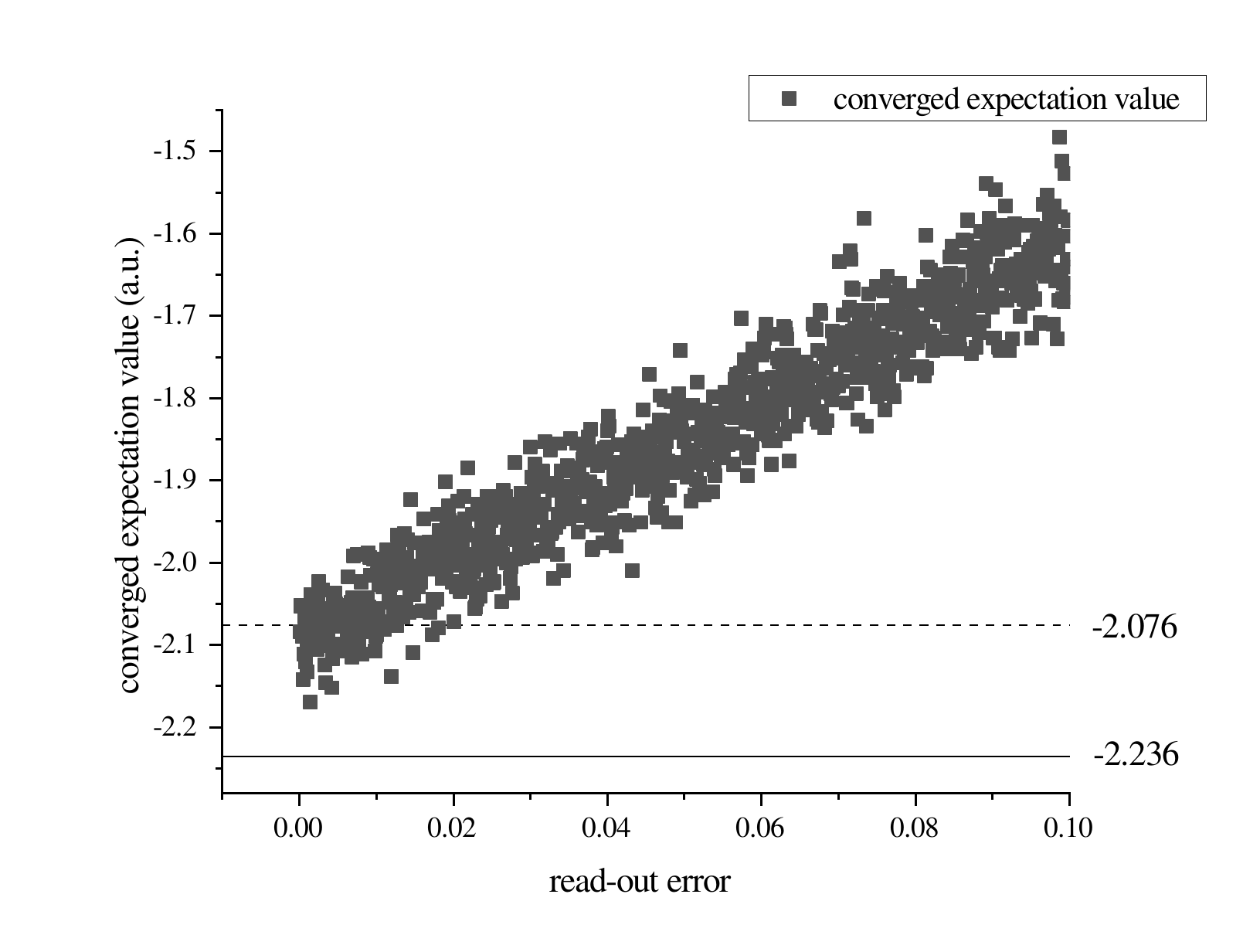}
\caption{\label{result_readout_error} The relation between the converged expectation value and read-out error.}
\end{figure}
To study the read-out error of our method based on Pauli term measurements, we simulate the VQE iterations with different read-out errors in Qiskit \cite{Qiskit}. The iteration times, initial guess, learning rate, and gate fidelity are set near the experiment. The final converged expectation value (and the distance from the true value) is found to have a positive correlation with the read-out error as illustrated in Fig. \ref{result_readout_error}. Since our converged minimum is -2.076, we can infer from the figure that the read-out error of our method is approximately at the magnitude of 0.01.

%
\bibliographystyle{IEEEtran}
\bibliography{output}

\end{document}